\documentstyle[12pt,epsf]{article}

\newcommand{\sect}[1]{\setcounter{equation}{0}\section{#1}}

\begin{document}

\title{Pair Creation of Dilaton Black Holes \\ in Extended Inflation}
\author{{\sc Raphael Bousso}\thanks{\it R.Bousso@damtp.cam.ac.uk}
      \\[1 ex] {\it Department of Applied Mathematics and}
      \\ {\it Theoretical Physics}
      \\ {\it University of Cambridge}
      \\ {\it Silver Street, Cambridge CB3 9EW}
       }
\date{DAMTP/R-96/42 \\[1ex] September 1996}

\maketitle

\begin{abstract}
  
  Dilatonic Charged Nariai instantons mediate the nucleation of black
  hole pairs during extended chaotic inflation. Depending on the
  dilaton and inflaton fields, the black holes are described by one of
  two approximations in the Lorentzian regime. For each case we find
  Euclidean solutions that satisfy the no boundary proposal. The
  complex initial values of the dilaton and inflaton are determined,
  and the pair creation rate is calculated from the Euclidean action.
  Similar to standard inflation, black holes are abundantly produced
  near the Planck boundary, but highly suppressed later on.  An
  unusual feature we find is that the earlier in inflation that the
  dilatonic black holes are created, the more highly charged they can
  be.

\end{abstract}

\pagebreak

\sect{Introduction}

If nature is described by a higher-dimensional theory such as string
theory, a dilaton field $\phi$ must be included in the low energy
action. This can alter the properties of black hole solutions
significantly. In standard Einstein-Maxwell theory with a cosmological
constant $\Lambda$, there is a family of charged non-rotating static
black hole solutions, the Reissner-Nordstr\"om-de~Sitter (RNdS)
solutions.  In dilatonic theories, a natural generalisation of a
cosmological constant term is a Liouville potential,
\begin{equation}
V(\phi) = \Lambda e^{2b\phi}.
\end{equation}
Poletti and Wiltshire showed that all RNdS solutions that approach
de~Sitter space asymptotically have no analogue in dilatonic theories
with a Liouville potential~\cite{PolWil94}. The solutions of highest
mass for given charge, however, have a different asymptotic behaviour
from all other RNdS solutions. These ``Charged Nariai'' solutions are
the product of 1+1-dimensional de~Sitter space with a round
two-sphere. They can be generalised to include a dilaton, which then
has a constant value~\cite{Bou96}. The Charged Nariai black holes
admit a smooth Euclidean section (an instanton), which is given by the
topological product of two round two-spheres, not necessarily of the
same radius.  Therefore these black holes can be spontaneously
created, if an appropriate background spacetime is available.

In the non-dilatonic case, this background is de~Sitter space. The
cosmological pair creation of non-dilatonic black holes is well
understood~\cite{MelMos89,Rom92,ManRos95}. One finds that pair
creation is highly suppressed unless the cosmological constant is on
the order of the Planck value. Since the universe does not have such a
large cosmological constant, an alternative background was used in
Refs.~\cite{BouHaw95,BouHaw96}: a chaotic inflationary universe. In
this case the potential of an inflaton field $\sigma$ acts as an
effective cosmological constant~\cite{Lin83,Haw84}:
\begin{equation}
V(\sigma) = m^2 \sigma^2.
\end{equation}
(For definiteness, we use a massive scalar field, but none of our
results depend qualitatively on this choice.) The inflationary
universe looks very much like de~Sitter space, except that the
cosmological constant is not fixed, but decreases slowly. As a
consequence, the Euclidean solutions must be chosen slightly complex,
so that the minisuperspace variables are perfectly real in the
Lorentzian regime~\cite{BouHaw95}. Black holes can be pair created
abundantly near the Planck era, when the effective cosmological
constant is large, but they become highly suppressed later on.  Most
neutral black holes evaporate before inflation ends, and magnetically
charged black holes, which are topologically conserved, are strongly
diluted by the inflationary expansion~\cite{BouHaw96}.  Thus there is
no conflict with present day observations.

In the dilatonic case, on the other hand, the Charged Nariai
instantons do not readily correspond to a pair creation process, since
there is no static de~Sitter solution that could act as a background.
Even if the dilaton is allowed to become time-dependent, it is pushed
towards negative infinity by the fixed cosmological constant. An
effective cosmological constant
\begin{equation}
V(\phi,\sigma) = e^{2b\phi}\, m^2 \sigma^2,
\end{equation}
however, is large only for a finite time. Therefore, an inflationary
background is still available; the dilaton will decrease only until
inflation ends. This extended model of chaotic inflation was
introduced by
Linde~\cite{Lin90} and has received considerable attention~\cite%
{BerMae91,DerGun92,Lid92,Gar94,GarLin94,GarLin95,GarWan95,GarWan95a}.
In this paper we investigate the pair creation of dilatonic black
holes during extended chaotic inflation through Charged Nariai
instantons. It will be interesting to see how the process differs from
pair creation during standard inflation. But what provides a more
fundamental motivation for this work is that inflation is the simplest
background on which cosmological dilaton black holes can be pair
created at all.

The paper is organised as follows. The theories we consider are
presented in Sec.~\ref{sec-dil-lag}. In Sec.~\ref{sec-no-bh} we make a
minisuperspace ansatz for the background dilatonic inflationary
universe. By choosing appropriate complex initial values, we find
solutions that satisfy the no boundary proposal at the origin of
Eudlidean time, and that are exactly real in the Lorentzian region. We
determine some observational constraints on the parameters of the
solution space. In Sec.~\ref{sec-bh} we go through the same procedure
for an inflationary universe containing dilatonic Charged Nariai black
holes. We find two approximate solutions, which are valid in different
regions of the dilaton-inflaton phase space. The first corresponds to
a sequence of the kind of fixed-$\Lambda$-solutions found in
Ref.~\cite{Bou96}. In the second, the black hole charge is negligible
and the dilaton evolves as if no Maxwell term were present. In
Sec.~\ref{sec-pc} we analyse the pair creation process during
inflation. We consider black holes created with a given charge at a
given point in inflation, and show that they are all attracted by one
of the two approximate solutions.  We calculate the Euclidean action
and pair creation rate in all cases, and show that pair creation is
suppressed except near the Planck boundary. We show that unlike in
standard inflation, the black holes created earliest have the highest
charge. A summary is provided in Sec.~\ref{sec-sum}.

\sect{Dilaton Lagrangian} \label{sec-dil-lag}

In the string frame, the gravitational Lagrangian with a dilaton and a massive
inflaton field is given by:
\begin{equation}
  L_{\rm string} = (-\bar{g})^{1/2} \left[ e^{-b\phi} \bar{R} - 2
  e^{-b\phi}\, \bar{g}^{\mu\nu} (\partial_\mu \phi) (\partial_\nu
  \phi) - 2 \bar{g}^{\mu\nu} (\partial_\mu \sigma) (\partial_\nu
  \sigma) - 2 m^2 \sigma^2 \right].
 \label{eq-string-lag}
\end{equation}
After a conformal transformation
\begin{equation}
g^{\mu\nu} \equiv e^{-b\phi}\, \bar{g}^{\mu\nu},
\end{equation}
the Lagrangian takes the Einstein-Hilbert form, but the dilaton field
now couples to the inflaton:
\begin{equation}
  L_{\rm Einstein} = (-g)^{1/2} \left[ R - 2 g^{\mu\nu} (\partial_\mu
  \phi) (\partial_\nu \phi) - 2 e^{b\phi}\, g^{\mu\nu} (\partial_\mu
  \sigma) (\partial_\nu \sigma) - 2 e^{2b\phi}\, m^2 \sigma^2 \right].
 \label{eq-einstein-lag}
\end{equation}
Throughout the paper we shall work in this frame, the Einstein frame.
In order to consider charged black holes, a Maxwell term must be added
to the Lagrangian:
\begin{equation}
L = L_{\rm Einstein} + (-g)^{1/2} \left[ - e^{-2a\phi} F^2 \right],
 \label{eq-full-lag}
\end{equation}
where
\begin{equation}
F^2 = F_{\mu\nu} F^{\mu\nu}.
\end{equation}

This Lagrangian is invariant under the transformation
\begin{equation}
  a \rightarrow -a,\ b \rightarrow -b,\ \phi \rightarrow -\phi.
\end{equation}
We fix a gauge by choosing $a \geq 0$. In order to obtain slow-roll
inflation, we must assume
\begin{equation}
|b| \ll 1.
\end{equation}
It was shown in Ref.~\cite{Bou96} that Nariai-type black hole
solutions of this theory exist only for $0<|b|<a$. They are
magnetically (electrically) charged for positive (negative) $b$. We
are interested mainly in magnetically charged black holes, since they
cannot evaporate. Thus we shall take
\begin{equation}
0 < b < a.
\end{equation}
It is easy to adapt our results to the electric case.

\sect{Inflation Without Black Holes} \label{sec-no-bh}

\subsection{Ansatz}

We first consider an inflating universe without black holes.  We make
the metric ansatz corresponding to a spatially closed
Friedmann-Robertson-Walker universe. Its spacelike sections are
three-spheres of radius $\alpha$:
\begin{equation}
ds^2 = N^2 d\tau^2 + \alpha(\tau)^2 d\Omega_3^2,
\end{equation}
where $N$ is the lapse function, $\tau$ is the Euclidean time
variable, and $d\Omega_3$ is the metric on the unit three-sphere.
From the Lagrangian, Eq.~(\ref{eq-einstein-lag}) we obtain the
Euclidean action
\begin{eqnarray}
  I_{S^3} & = & - \frac{3\pi}{4} \int N d\tau \left[ \alpha +
  \frac{\alpha \dot{\alpha}^2}{N^2} \right. \nonumber \\ & & \left.
  \mbox{} - \frac{1}{3} \alpha^3 \left( \frac{\dot{\phi}^2}{N^2} +
  e^{b\phi} \frac{\dot{\sigma}^2}{N^2} + e^{2b\phi}\, m^2 \sigma^2
\right) \right].
 \label{eq-action-frw}
\end{eqnarray}
Variation with respect to $\alpha$, $\sigma$, $\phi$ and $N$ yields
the equations of motion and the Hamiltonian constraint:
\begin{eqnarray}
  \frac{\ddot{\alpha}}{\alpha} + \frac{2}{3} \dot{\phi}^2 +
  \frac{2}{3} e^{b\phi}\, \dot{\sigma}^2 + \frac{1}{3} e^{2b\phi}\,
  m^2 \sigma^2 &
  = & 0,
\label{eq-m-frw-alpha} \\
\ddot{\sigma} + 3 \frac{\dot{\alpha}}{\alpha} \dot{\sigma}  + b
\dot{\phi} \dot{\sigma} - e^{b\phi}\, m^2 \sigma & = & 0,
\label{eq-m-frw-sig} \\
\ddot{\phi} + 3 \frac{\dot{\alpha}}{\alpha} \dot{\phi} - \frac{b}{2}
e^{b\phi}\, \dot{\sigma}^2 - b e^{2b\phi}\, m^2 \sigma^2 & = & 0,
\label{eq-m-frw-phi} \\
\frac{1}{\alpha^2} - \frac{\dot{\alpha}^2}{\alpha^2} + \frac{1}{3}
\dot{\phi}^2 + \frac{1}{3} e^{b\phi}\, \dot{\sigma}^2 - \frac{1}{3}
e^{2b\phi}\, m^2 \sigma^2 & = & 0,
\label{eq-m-frw-N}
\end{eqnarray}
in the gauge $N=1$, where an overdot denotes differentiation with
respect to $\tau$.

\subsection{Solutions}

It is convenient to define
\begin{eqnarray} 
H^2 & = & \mbox{Re}\left( \frac{1}{3} e^{2b\phi}\, m^2 \sigma^2 \right),
\\
\theta & = & e^{-b\phi}.
\end{eqnarray}
We must find solutions that satisfy the no boundary proposal:
\begin{equation} 
\begin{array}{l}
  \alpha=0,\ \dot{\alpha}=1,\ \sigma=\sigma_0,\ \dot{\sigma}=0,\\
  \theta=\theta_0,\ \dot{\theta}=0\ \ \mbox{for}\ \ \tau=0.
\end{array}
\label{eq-nbp-frw}
\end{equation}
We shall assume that for both the dilaton and inflaton, the initial
real value of the field is much greater than the initial imaginary
value. An approximate solution near the origin, for $|H_0 \tau| <
O(1)$, is given by
\begin{eqnarray}
\sigma_{\cal I}(\tau) & = & \sigma_{0},
\label{eq-frw-sigmaI} \\
\theta_{\cal I}(\tau) & = & \theta_{0},
\label{eq-frw-thetaI} \\
\alpha_{\cal I}(\tau) & = & H_0^{-1} \sin H_0 \tau.
\label{eq-frw-tauI}
\end{eqnarray}
This ``inner'' approximation satisfies the no boundary proposal.

We define a Lorentzian time variable $t$ by
\begin{equation}
\tau = \frac{\pi}{2H_0} + it,
\end{equation}
For $H_0 t > O(1)$ an approximate solution is
\begin{eqnarray}
\sigma(t) & = & \bar{\sigma}_0 - \frac{m}{\sqrt{3}} t,
\label{eq-frw-sig} \\
\theta(t) & = & \bar{\theta}_0 + b^2 \frac{m}{\sqrt{3}} \int_0^t
\sigma\, dt',
\label{eq-frw-theta} \\
\alpha(t) & = & H^{-1} \cosh \int_0^t H dt'.
\label{eq-frw-alpha}
\end{eqnarray}
The constants $\bar{\sigma}_0$ and $\bar{\theta}_0$ denote the real
parts of the values of the two fields in the initial Euclidean region.
This ``outer'' approximation describes an inflating Lorentzian
universe, as long as the condition $|\dot{H}| \ll H^2$ is satisfied,
or equivalently, if
\begin{equation}
\frac{\sigma^2}{\theta} \gg 1.
\label{eq-cond-infl}
\end{equation}
Inflation ends when $|\dot{H}| \approx H^2$.

As in ordinary chaotic inflation~\cite{BouHaw95}, the fields must
start out with a non-zero imaginary part, in order to be perfectly
real during the Lorentzian inflationary era, i.e. for $\tau^{\rm Re} =
\frac{\pi}{2H_0}$. This imaginary part can be determined as follows.
Because of the initial conditions imposed by the no boundary proposal,
Eq.~(\ref{eq-nbp-frw}), both $\theta$ and $\sigma$ will be even
functions of $\tau$. Therefore the imaginary parts of both fields will
be constant along the imaginary $\tau$-axis, and will be equal to the
initial imaginary values:
\begin{equation}
  \theta^{\rm Im} = \theta_{0}^{\rm Im},\ \ \sigma^{\rm Im} =
  \sigma_{0}^{\rm Im}.
\label{eq-in-frw}
\end{equation}
In the outer approximation the imaginary parts of the two fields are
also constant along the imaginary $\tau$-axis, where they must have
the values.
\begin{equation} 
\theta^{\rm Im} = \frac{\pi}{2} b^2 \theta_0,\ \
\sigma^{\rm Im} = - \frac{\pi}{2} \frac{\theta_0}{\sigma_0},
\label{eq-out-frw}
\end{equation}
in order to vanish on the line $\tau^{\rm Re} = \frac{\pi}{2H_0}$.
This follows from Eqs.~(\ref{eq-frw-sig}) and (\ref{eq-frw-theta}),
where we have neglected the $t^2$-term.  Thus it does not matter
exactly at which point on the imaginary $\tau$-axis we match the
approximations; if we match at some $\tau^{\rm Im} \approx
O(H_0^{-1})$, then Eqs.~(\ref{eq-in-frw}) and (\ref{eq-out-frw})
determine $\theta_{0}^{\rm Im}$ and $\sigma_{0}^{\rm Im}$.

\subsection{Constraints}

Certain constraints are placed on the solution parameters by the
requirement that inflation must produce the density fluctuations we
observe in today's universe. For definiteness we shall assume that
inflation begins at the Planck boundary,
\begin{equation}
  e^{2b\phi_{\rm i}}\, m^2 \sigma_{\rm i}^2 = 1,
\end{equation}
and ends, by Eq.~(\ref{eq-cond-infl}), when
\begin{equation}
  \frac{\sigma_{\rm e}^2}{\theta_{\rm e}} = 1.
\end{equation}
From Eqs.~(\ref{eq-frw-sig}) and (\ref{eq-frw-theta}) it follows that
\begin{equation}
\sigma^2 + \frac{2}{b^2} \theta = \mbox{const} \equiv \zeta^2.
\label{eq-zeta}
\end{equation}
If we assume that inflation starts in the Planck era, the constant
$\zeta^2$ is the only parameter of the solution space.
Eq.~(\ref{eq-zeta}) allows us to relate the initial and final values
of the inflaton field:
\begin{equation}
  \sigma_{\rm e}^2 \left( 1 + \frac{2}{b^2} \right) = \zeta^2 =
  \sigma_{\rm i}^2 + \frac{2}{b^2} m \sigma_{\rm i}.
\label{eq-zeta2}
\end{equation}

The number of $e$-foldings by which the spatial size of the universe
grows during inflation is given by
\begin{equation}
N = \int_{\sigma_{\rm i}}^{\sigma_{\rm e}} H dt.
\end{equation}
This can be integrated using Eq.~(\ref{eq-zeta}):
\begin{eqnarray}
  N & = & \frac{1}{b^2} \int_{\sigma_{\rm i}}^{\sigma_{\rm e}}
  \frac{-2\sigma d\sigma}{\zeta^2 - \sigma^2} \nonumber \\ & = &
  \frac{1}{b^2} \left[ \ln ( \zeta^2 - \sigma^2 ) \right]_{\sigma_{\rm
      i}}^{\sigma_{\rm e}}.
\end{eqnarray}
Using Eq.~(\ref{eq-zeta2}) to eliminate $\zeta$ and $\sigma_{\rm i}$, we
find
\begin{eqnarray}
  N & = & \frac{1}{b^2} \ln \left( \frac{\sigma_{\rm
      e}^2}{m \sigma_{\rm i}} \right) \nonumber \\ & = & - \frac{1}{2}
  + \frac{1}{b^2} \ln \left( \frac{Y}{ -1 +
    \sqrt{1+2Y}} \right),
\end{eqnarray}
where
\begin{equation}
  Y \equiv \frac{\sigma_{\rm e}^2 b^2}{m^2}
  \left( 1 + \frac{b^2}{2} \right).
\end{equation}
The minimum number of $e$-foldings of inflation needed to account for
the large scale homogeneity of the observable universe is $N_{\rm min}
\approx 60$. This places a lower bound on $Y$:
\begin{equation}
Y > 2 X (X-1),
\end{equation}
where
\begin{equation}
  X \equiv \exp \left[ b^2 \left( N_{\rm min} + \frac{1}{2} \right)
  \right].
\end{equation}
Normally $b$ will be so small that we can approximate
\begin{equation}
X \approx 1 + b^2 N_{\rm min}.
\end{equation}
We then find the following condition for sufficient inflation:
\begin{equation}
\frac{\sigma_{\rm e}^2}{m^2} > 2 N_{\rm min}.
\end{equation}

The density perturbation on the horizon scale of the observable
universe is given by~\cite{COBE}:
\begin{equation}
\frac{\delta \rho}{\rho} \approx 5 \times 10^{-5}.
\end{equation}
The perturbations generated by the inflationary model we use have been
computed elsewhere~\cite{BerMae91,DerGun92,GarLin94}; here we only
quote the result:
\begin{equation}
\frac{\delta \rho}{\rho} \approx \frac{m}{\theta}.
\end{equation}
This leads to the constraint
\begin{equation}
  \frac{m}{\theta_{\rm e}} = \frac{m}{\sigma_{\rm e}^2} \approx 5
  \times 10^{-5}.
\end{equation}

\sect{Inflation With Black Holes} \label{sec-bh}

\subsection{Ansatz}

We now consider black hole solutions to the inflationary model of
Eq.~(\ref{eq-full-lag}). The topology of the spatial sections of such
solutions is $S^1 \times S^2$. In general, the size of the two-spheres
can vary along the one-sphere. In the case where one has a fixed
cosmological constant and no dilaton, such solutions approach
de~Sitter space beyond the cosmological horizon. It was shown by
Poletti and Wiltshire~\cite{PolWil94} that these solutions possess no
analogues in dilatonic theories with a Liouville potential, because of
the absence of a de~Sitter-like solution that could act as a
background. The only exception are solutions in which the two-sphere
radius is constant along the $S^1$. They correspond to black holes of
maximal mass at a given charge, and are called Charged Nariai
solutions. They do not approach de~Sitter space asymptotically; thus
they possess dilatonic analogues.

For a Liouville potential ($\Lambda e^{2b\phi}$) the dilatonic Charged
Nariai solutions are given in Ref.~\cite{Bou96}, where it is shown
that in some respects they differ quite radically from their
non-dilatonic counterparts. Since they admit a regular Euclidean
section, they can mediate black hole pair creation on a suitable
de~Sitter-like background. For fixed $\Lambda$ there is no such
background, because the cosmological constant pushes the dilaton
toward negative infinity.  The action, Eq.~(\ref{eq-einstein-lag}),
however, contains a Liouville potential with an effective cosmological
constant $\Lambda_{\rm eff} = m^2 \sigma^2$. The dilaton will only
decrease as long as $\Lambda_{\rm eff}$ is large; therefore extended
chaotic inflation provides an appropriate background for the pair
creation of the dilatonic Charged Nariai solutions of
Ref.~\cite{Bou96}. They need only be slightly modified to take the
time dependence of the cosmological constant into account.

Since the spacelike sections will be the direct product of a
one-sphere and a round two-sphere, we make the metric ansatz
\begin{equation}
ds^2 = N^2 d\tau^2 + \alpha(\tau)^2 d\xi^2 + \beta(\tau)^2
d\Omega_2^2,
\end{equation}
where $\xi$ has the period $2\pi$, and $d\Omega_2^2 = d\theta^2 +
\sin^2\! \theta\, d\varphi^2$. The equation of motion due to the
Maxwell term in Eq.~(\ref{eq-full-lag}),
\begin{equation}
0 = \nabla_{\mu} \left( e^{-2a\phi} F^{\mu\nu} \right),
  \label{eq-dil-motion-F}
\end{equation}
can be integrated to yield
\begin{equation}
F = Q \sin \theta \, d\theta \wedge d\varphi
\end{equation}
in the case of magnetically charged black holes. Thus we have $F^2 =
2 Q^2 / \beta^4$, and we can obtain the Euclidean minisuperspace
action from Eq.~(\ref{eq-full-lag}):
\begin{eqnarray} 
  I_{S^1 \times S^2} & = & -\pi \int N d\tau \left[ \alpha +
  \frac{\alpha \dot{\beta}^2}{N^2} + \frac{2 \dot{\alpha} \dot{\beta}
    \beta}{N^2} \right. \nonumber \\
& & \left. \mbox{} - \alpha \beta^2 \left(
  \frac{\dot{\phi}^2}{N^2} + e^{b\phi} \frac{\dot{\sigma}^2}{N^2} +
  e^{2b\phi} m^2 \sigma^2 + e^{-2a\phi} \frac{Q^2}{\beta^4} \right)
\right] \nonumber \\
& & \mbox{} + \pi \left[ -\dot{\alpha} \beta^2 - 2 \alpha
\beta \dot{\beta} \right]_{\tau=0} .
 \label{eq-action-bh}
\end{eqnarray} 
Variation with respect to $\alpha$, $\beta$, $\sigma$, $\phi$ and $N$
yields the equations of motion and the Hamiltonian constraint:
\begin{eqnarray}
\frac{\ddot{\beta}}{\beta} - \frac{\dot{\alpha} \dot{\beta}}{\alpha
  \beta} + \dot{\phi}^2 + e^{b\phi}\, \dot{\sigma}^2 & = & 0,
\label{eq-m-bh-alpha} \\
\frac{\ddot{\beta}}{\beta} + \frac{\dot{\alpha} \dot{\beta}}{\alpha
  \beta} + \frac{\ddot{\alpha}}{\alpha} + \dot{\phi}^2 + e^{b\phi}\,
\dot{\sigma}^2 + e^{2b\phi}\, m^2 \sigma^2 - e^{-2a\phi}\,
\frac{Q^2}{\beta^4} & = & 0,
\label{eq-m-bh-beta} \\
\ddot{\sigma} + \left( \frac{\dot{\alpha}}{\alpha} + 2
\frac{\dot{\beta}}{\beta} \right) \dot{\sigma} + b
\dot{\phi} \dot{\sigma} - e^{b\phi}\, m^2 \sigma & = & 0,
\label{eq-m-bh-sig} \\
\ddot{\phi} + \left( \frac{\dot{\alpha}}{\alpha} + 2
\frac{\dot{\beta}}{\beta} \right) \dot{\phi} - \frac{b}{2}
e^{b\phi}\, \dot{\sigma}^2 - b e^{2b\phi}\, m^2 \sigma^2 +
a e^{-2a\phi}\, \frac{Q^2}{\beta^4} & = & 0,
\label{eq-m-bh-phi} \\
\frac{1}{\beta^2} - \frac{\dot{\beta}^2}{\beta^2} - 2
\frac{\dot{\alpha} \dot{\beta}}{\alpha \beta} + \dot{\phi}^2 +
e^{b\phi}\, \dot{\sigma}^2 - e^{2b\phi}\, m^2 \sigma^2 - e^{-2a\phi}\,
\frac{Q^2}{\beta^4} & = & 0,
\label{eq-m-bh-N}
\end{eqnarray}
in the gauge $N=1$.

We define
\begin{eqnarray} 
\tilde{H}^2 & = & \mbox{Re}\left( e^{2b\phi}\, m^2 \sigma^2 \right),
\\
\tilde{g} & = & \frac{b}{a}.
\end{eqnarray}
The black hole solutions must satisfy the no boundary proposal:
\begin{equation} 
\begin{array}{l}
  \alpha=0,\ \dot{\alpha}=1,\ \sigma=\sigma_0,\ \dot{\sigma}=0,\\
  \beta=\beta_0,\ \dot{\beta}=0,\
  \theta=\theta_0,\ \dot{\theta}=0\ \ \mbox{for}\ \ \tau=0.
\end{array}
\label{eq-nbp-bh}
\end{equation}
Again it is assumed that the initial real values of both fields are
much greater than the initial imaginary values, and that we are
sufficiently far from the end of inflation, i.e.\ $\sigma^2 / \theta
\gg 1$.

In the dilatonic Charged Nariai solutions with a fixed cosmological
constant, the dilaton is fixed at the value
\begin{equation}
  \phi_{\rm eq} = \frac{1}{2a(1-\tilde{g})} \ln \left[
  \frac{(1+\tilde{g})^2}{\tilde{g}} Q^2 \Lambda \right],
\end{equation}
for which the cosmological and Maxwell terms in the dilaton equation
of motion cancel out~\cite{Bou96}. But now we are dealing with an
effective cosmological constant, which decreases slowly. Therefore
$\phi_{\rm eq}$ will also change over time:
\begin{equation}
  \phi_{\rm eq}(\sigma) = \frac{1}{2a(1-\tilde{g})} \ln \left[
  \frac{(1+\tilde{g})^2}{\tilde{g}} Q^2 m^2 \sigma^2 \right].
\label{eq-phieq}
\end{equation}
There will thus be two cases. One possibility is that the dilaton
evolution follows that of $\phi_{\rm eq}$, so that the cosmological
and Maxwell terms remain opposite and equal as the inflaton field
decreases. In this case the time evolution of the inflationary black
hole solution can be approximated as a sequence of static solutions
with $\Lambda = \Lambda_{\rm eff}$. The other possibility is that
$\phi_{\rm eq}$ changes too fast for $\phi$ to follow.
Correspondingly, we shall give two different approximations for the
black hole solutions, valid in different regions of the
$(\phi,\sigma)$ phase space. Each will split into an inner and outer
approximation.

\subsection{Pseudo-static Approximation}

First consider the ``pseudo-static'' case where $\phi \approx
\phi_{\rm eq}$. Then the inner approximation, valid for $|\tilde{H}_0
(1-\tilde{g})^{1/2}\, \tau| < O(1)$, is given by
\begin{eqnarray}
\sigma_{\cal I}(\tau) & = & \sigma_{0},
\label{eq-bh-sigI} \\
\phi_{\cal I}(\tau) & = & \phi_{\rm eq}(\sigma_0),
\label{eq-bh-phiI} \\
\alpha_{\cal I}(\tau) & = & \tilde{H}_0^{-1} (1-\tilde{g})^{-1/2} \sin
\tilde{H}_0 (1-\tilde{g})^{1/2}\, \tau,
\label{eq-bh-alphaI} \\
  \beta_{\cal I}(\tau) & = & \tilde{H}_0^{-1} (1+\tilde{g})^{-1/2}.
\label{eq-bh-betaI}
\end{eqnarray}
With the Lorentzian time variable t defined by
\begin{equation}
\tau = \frac{\pi}{2\tilde{H}_0 (1-\tilde{g})^{1/2}} + it,
\end{equation}
the outer approximation, valid for $\tilde{H}_0 (1-\tilde{g})^{1/2}\,
t > O(1)$, is given by
\begin{eqnarray}
\sigma(t) & = & \bar{\sigma}_0 - (1-\tilde{g})^{-1/2} m t,
\label{eq-bh-sig} \\
\phi(t) & = & \phi_{\rm eq}(\sigma),
\label{eq-bh-phi} \\
\alpha(t) & = & \tilde{H}^{-1} (1-\tilde{g})^{-1/2} \cosh \int_0^t
\tilde{H} (1-\tilde{g})^{1/2} dt',
\label{eq-bh-alpha} \\
\beta(t) & = & \tilde{H}^{-1} (1+\tilde{g})^{-1/2}.
\label{eq-bh-beta}
\end{eqnarray}
This describes an inflating universe containing a pair of dilatonic
black holes of charge $Q$.  The imaginary initial value of $\sigma$
can be determined in the way described in the previous section:
\begin{equation}
\sigma_0^{\rm Im} = - (1-\tilde{g})^{-1/2} \frac{\pi}{2}
\frac{\theta_0}{\sigma_0}.
\end{equation}
Then the values of $\phi_0^{\rm Im}$ and $\beta_0^{\rm Im}$ follow
from Eqs.~(\ref{eq-bh-phiI}) and (\ref{eq-bh-betaI}).

In this approximation $\phi$ decreases along with $\phi_{\rm eq}$. It
is pushed down by the effective potential due to the cosmological and
Maxwell terms (the last two terms on the left hand side of
Eq.~(\ref{eq-m-bh-phi})): if $\phi > \phi_{\rm eq}$, the Maxwell term
will be smaller than the cosmological term, and one obtains
$\frac{d\phi}{dt} < 0$. However, the rate of decrease of $\phi$ has a
limit corresponding to the dropping of the Maxwell term:
\begin{equation}
\left| \frac{d\phi}{dt} \right| < \min \left\{
(1-\tilde{g})^{-1/2}\, b m \sigma e^{b\phi},
\frac{1}{2} (1- \tilde{g})^{3/2} b m \sigma^3 e^{2b\phi} \right\}
\end{equation}
(The two different bounds come from neglecting either one or the other
of the two terms in the prefactor of $ \dot{\phi} $ in
Eq.~(\ref{eq-m-bh-phi}).) Thus $\phi$ can only keep pace with
$\phi_{\rm eq}$ as long as
\begin{equation}
  \left| \frac{d\phi_{\rm eq}}{dt} \right| = \frac{1}{a}
    (1-\tilde{g})^{-3/2}\, \frac{m}{\sigma}
\end{equation}
remains within this limit. This gives us a condition for the validity
of the pseudo-static approximation:
\begin{equation}
\sigma^2 e^{b\phi} \gg \max \left\{
\frac{1}{a^2 \tilde{g} (1-\tilde{g})},
\left[ \frac{1}{a^2 \tilde{g} (1- \tilde{g} )^3} \right]^{1/2}
\right\}.
\label{eq-high}
\end{equation}
We call the region of phase space where this equation holds the {\em
  high field regime}.

\subsection{Neutral Approximation}

Since $\sigma^2$ and $e^{b\phi}$ both decrease in the high field
regime, the high field condition, Eq.~(\ref{eq-high}), will eventually
cease to hold. We enter the {\em low field regime}, where $\phi$ will
start to trail behind $\phi_{\rm eq}$.  Then the Maxwell term in
Eq.~(\ref{eq-m-bh-phi}) will become smaller than the cosmological term
and can soon be neglected altogether. Since $b/a<1$ the Maxwell terms
in the other equations of motion can be dropped as well.  In this
limit we can obtain another approximation, which resembles a neutral
black hole solution. This regime is either reached as a second stage
after a period of pseudo-static evolution, or it emerges from a
Euclidean instanton in a process of pair creation. In the latter case
it describes the nucleation of black holes in both the high and low
field regimes, as long as the black hole charge is small enough to
make the Maxwell term negligible. In order to include the Euclidean
section, we begin by giving the inner neutral approximation
($|\tilde{H}_0 \tau| < O(1)$):
\begin{eqnarray}
\sigma_{\cal I}(\tau) & = & \sigma_0,
\label{eq-neutral-sigI} \\
\theta_{\cal I}(\tau) & = & \theta_0,
\label{eq-neutral-phiI} \\
\alpha_{\cal I}(\tau) & = & \tilde{H}_0^{-1} \sin \tilde{H}_0 \tau,
\label{eq-neutral-alphaI} \\
\beta_{\cal I}(\tau)  & = & \tilde{H}_0^{-1}.
\label{eq-neutral-betaI}
\end{eqnarray}
With
\begin{equation}
\tau = \frac{\pi}{2\tilde{H}_0} + it,
\end{equation}
the outer neutral approximation ($\tilde{H}_0 t > O(1)$) is given by
\begin{eqnarray}
\sigma(t) & = & \bar{\sigma}_0 - m t,
\label{eq-neutral-sig} \\
\theta(t) & = & \bar{\theta}_0 + b^2 m \int_0^t \sigma\, dt',
\label{eq-neutral-theta} \\
\alpha(t) & = & \tilde{H}^{-1} \cosh \int_0^t \tilde{H} dt',
\label{eq-neutral-alpha} \\
\beta(t) & = & \tilde{H}^{-1}.
\label{eq-neutral-beta}
\end{eqnarray}
This describes an inflationary universe containing a dilatonic black
hole pair of negligible charge. As in the FRW solution, the constraint
\begin{equation}
\sigma^2 + \frac{2}{b^2} \theta = \mbox{const}
\label{eq-zeta3}
\end{equation}
is satisfied. The imaginary parts of the initial values of the
inflaton and dilaton are
\begin{equation}
\theta_0^{\rm Im} = \frac{\pi}{2} b^2 \theta_0,\ \
\sigma_0^{\rm Im} = - \frac{\pi}{2} \frac{\theta_0}{\sigma_0}.
\end{equation}
The value of $\beta_0^{\rm Im}$ follows from
Eq.~(\ref{eq-neutral-betaI}).

\sect{Black Hole Pair Creation} \label{sec-pc}

We have found instanton solutions for a inflationary universe with a
dilaton, and for a black hole pair immersed in this background. Thus
black holes can spontaneously appear during extended chaotic
inflation. The pair creation rate is given by
\begin{equation}
  \Gamma = \exp \left[ - \left( 2 I_{\rm bh}^{\rm Re} - 2 I_{\rm
    bg}^{\rm Re} \right) \right],
\label{eq-pcr}
\end{equation}
where $I_{\rm bh}^{\rm Re}$ and $I_{\rm bg}^{\rm Re}$ are the real
parts of the Euclidean actions of the black hole and background
instantons. We shall discuss pair creation in the high and low field
regimes separately.

\subsection{High Field Regime}

\subsubsection*{Pseudo-static Creation}

\begin{figure}[htb]
  \hspace*{\fill} \vbox{\epsfxsize=.5\textwidth
  \epsfbox{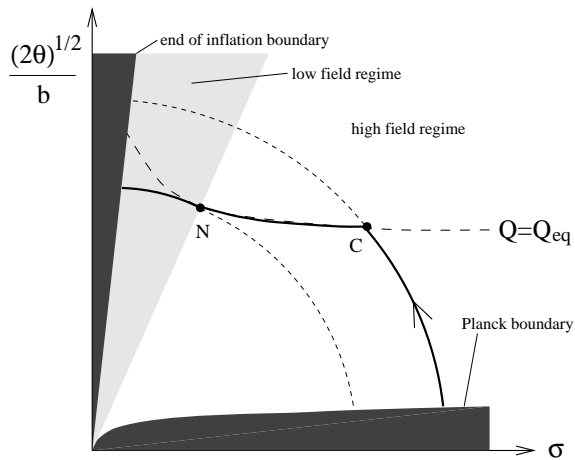}} \hspace*{\fill}
\caption[Creation of a dilaton black hole in the high field regime with
$Q=Q_{\rm eq}$]%
{\small\sl Creation of a dilaton black hole in the high field regime
  with $Q=Q_{\rm eq}$. The thick line indicates schematically the
  corresponding path in the dilaton-inflaton phase space. Inflation
  begins at the Planck boundary. Black hole pair creation takes place
  at {\rm C}. The transition from the pseudo-static to the neutral
  approximation takes place around {\rm N}.}
\label{fig-eci1}
\end{figure}

For the black hole solutions in the high field regime, we found the
pseudo-static approximation, Eqs.~(\ref{eq-bh-sigI}) --
(\ref{eq-bh-beta}).  If pseudo-static black holes are created via an
instanton, the black hole charge is entirely fixed by the background:
by rewriting Eq.~(\ref{eq-phieq}) the charge can be expressed as a
function of the dilaton and inflaton field:
\begin{equation}
  Q_{\rm eq} = \frac{\tilde{g}^{1/2}}{1+\tilde{g}}\,
  \frac{e^{a(1-\tilde{g})\phi}}{m \sigma}.
\label{eq-qeq}
\end{equation}
In fact black holes of different charge can be also be created, but we
leave this discussion until later. In order to become familiar with
the scenario, let us first consider the pair creation of a black hole
of charge $Q_{\rm eq}$ in the high field regime, where
Eq.~(\ref{eq-high}) holds, and where it can then evolve according to
the pseudo-static approximation. This process can be depicted in a
dilaton-inflaton phase space diagram (Fig.~\ref{fig-eci1}).
By Eq.~(\ref{eq-zeta}) the paths traced out by the fields during
inflation without a black hole lie on concentric circles around the
origin. By Eq.~(\ref{eq-qeq}), the lines of equal charge are
hyperbolae. We expect the inflationary universe to start out near the
Planck boundary and move counterclockwise along one of the circles.
After the pair creation process takes place, the black hole will be
described by a pseudo-static approximation. This means that the fields
depart from the circle and move instead on the hyperbola corresponding
to the black hole charge $Q_{\rm eq}$. Eventually the evolution moves
out of the high field and into the low field regime, where $\phi$ can
no longer follow $\phi_{\rm eq}$. The evolution will then be described
by the neutral approximation. Therefore the fields will leave the
hyperbola and, by Eq.~(\ref{eq-zeta3}), move once again on a circle,
smaller than the original one, until inflation ends.

By Eq.~(\ref{eq-qeq}),
\begin{equation}
  \frac{dQ_{\rm eq}}{dt} = \frac{1}{1+\tilde{g}}\, \left(
  \frac{\tilde{g}}{3} \right)^{1/2} \theta^{-1/\tilde{g}} \left[
  \frac{\theta}{\sigma^2} - a^2 \tilde{g} (1-\tilde{g}) \right].
\label{eq-dq}
\end{equation}
In the high field regime, the right hand side is negative, by
Eq.~(\ref{eq-high}). Therefore pseudo-static black holes have a higher
charge, the {\em earlier} in inflation they are created.  This feature
might seem surprising at first. For black holes pair created in
standard chaotic inflation, the maximum charge increases as the
cosmological constant runs down~\cite{BouHaw96}. This is because the
size of the black holes is inversely proportional to $H$. The black
holes created later will thus be larger and can support a higher
charge. The sharp contrast with the dilatonic case can be explained by
the variability of the Maxwell coupling $e^{-2a\phi}$. This factor
increases during inflation, overcompensating for the growth of the
black hole radius $\beta$ in the Maxwell term $e^{-2a\phi}\,
Q^2/\beta^4$. In the low field regime, the right hand side of
Eq.~(\ref{eq-dq}) is positive, and so the black hole charge will
increase.

Using the inner solution, Eqs.~(\ref{eq-frw-sigmaI}) --
(\ref{eq-frw-tauI}), we can calculate the real part of the Euclidean
background action, Eq.~(\ref{eq-action-frw}):
\begin{equation}
I_{S^3}^{\rm Re} = - \frac{\pi}{2 H_0^2}.
\label{eq-action-frw2}
\end{equation}
In the black hole case the Euclidean action comes entirely from the
$\tau=0$ term in Eq.~(\ref{eq-action-bh}) and is given by
\begin{equation}
  I_{S^1 \times S^2}^{\rm Re} = - \frac{\pi}{\tilde{H}_0^2
    (1+\tilde{g})} = - \frac{\pi}{3 H_0^2 (1+\tilde{g})}.
\label{eq-action-bh2}
\end{equation}
Using Eq.~(\ref{eq-pcr}), the pair creation rate for the pseudo-static
black holes can be determined:
\begin{equation}
  \Gamma_{\rm pseudo} = \exp \left( - \frac{\pi}{3 H_0^2}\, \frac{1 +
    3 \tilde{g}}{1+\tilde{g}} \right).
\label{eq-pcr-pseudo}
\end{equation}
The result is qualitatively the same as in the non-dilatonic
case~\cite{BouHaw96}. Since the exponent is negative, black holes are
suppressed relative to the background, as they should be. Near the
Planckian regime, where $H \approx 1$, the suppression is weak and
many black holes can be created. As the dilaton and inflaton fields
decrease, $H$ becomes much smaller than 1, and the pair creation will
be exponentially suppressed.

\subsubsection*{Other Charges}

\begin{figure}[htb]
  \hspace*{\fill} \vbox{\epsfxsize=.5\textwidth
  \epsfbox{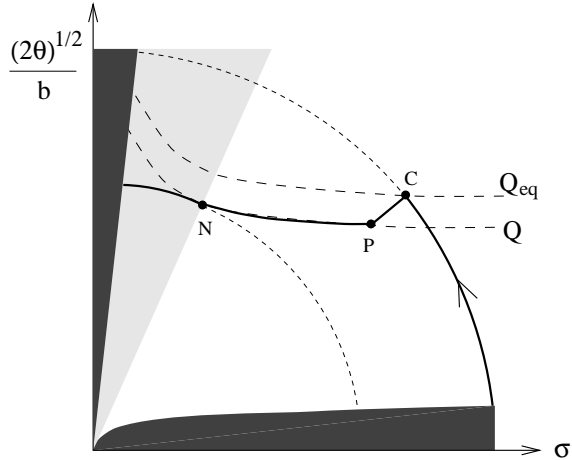}} \hspace*{\fill}
\caption[Creation of a dilaton black hole in the high field regime with
$Q>Q_{\rm eq}$]%
{\small\sl Creation of a dilaton black hole in the high field regime
  with $Q>Q_{\rm eq}$. After the black hole is created at {\rm C}, the
  dilaton increases to compensate for the high charge. Between {\rm P}
  and {\rm N} the evolution is described by a pseudo-static
  approximation and after {\rm N} by a neutral approximation.}
\label{fig-eci2}
\end{figure}

We will now discuss the case where the black holes are created in the
high field regime, but not with the charge given by
Eq.~(\ref{eq-qeq}). At least initially, they will therefore not be
described by the pseudo-static approximation. It is convenient to
define the variable $g'$, which takes the role of $\tilde{g}$ in the
initial Euclidean section:
\begin{equation}
\frac{g'}{(1+g')^2} = e^{-2(a-b)\phi}\, Q^2 m^2 \sigma^2.
\end{equation}
By Eqs.~(\ref{eq-m-bh-beta}) and (\ref{eq-m-bh-N}), the radii of the
Euclidean $S^2 \times S^2$ will be approximately $\tilde{H}_0^{-1}
(1-g')^{-1/2}$ and $\tilde{H}_0^{-1} (1+g')^{-1/2}$. Thus the charge
has an upper limit determined by the condition $g'<1$.  If
$g'>g$ (or equivalently, $Q>Q_{\rm eq}$) the Maxwell term is
larger than the cosmological term in the dilaton equation of motion.
Thus the dilaton will increase as the inflaton decreases, and
eventually it will reach the value $\phi_{\rm eq}$ given by
Eq.~(\ref{eq-phieq}). The evolution of the black hole will then settle
down to the pseudo-static approximation (or the neutral approximation,
if Eq.~(\ref{eq-high}) no longer holds by the time that $\phi =
\phi_{\rm eq}$), as shown in Fig.~\ref{fig-eci2}.

\begin{figure}[htb]
  \hspace*{\fill} \vbox{\epsfxsize=.5\textwidth
  \epsfbox{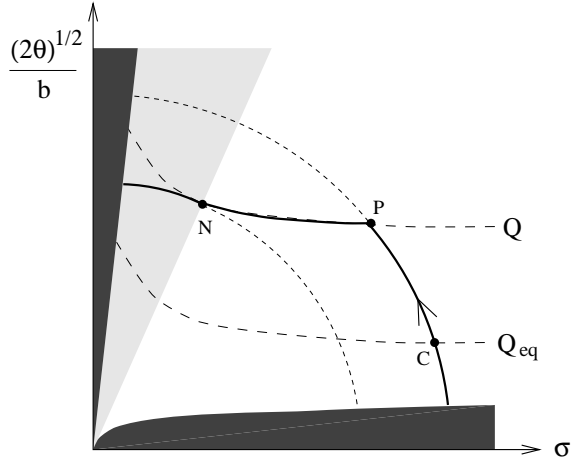}} \hspace*{\fill}
\caption[Creation of a dilaton black hole in the high field regime with
$Q<Q_{\rm eq}$]%
{\small\sl Creation of a dilaton black hole in the high field regime
  with $Q<Q_{\rm eq}$. A black hole created at {\rm C} is initially
  described by a neutral approximation. It enters a pseudo-static
  phase at {\rm P} and another neutral phase at {\rm N}.}
\label{fig-eci3}
\end{figure}

\begin{figure}[htb]
  \hspace*{\fill} \vbox{\epsfxsize=.5\textwidth
  \epsfbox{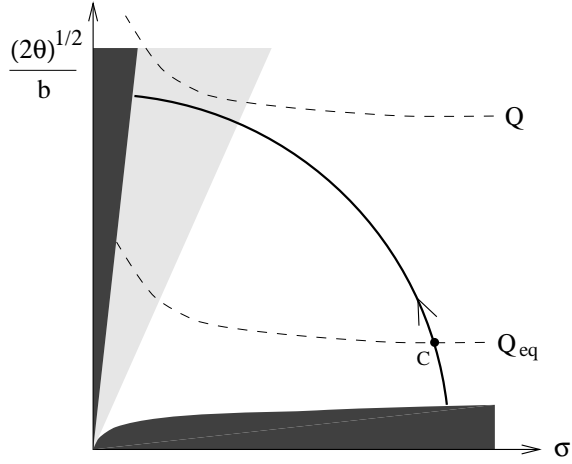}} \hspace*{\fill}
\caption[Creation of a dilaton black hole in the high field regime with
$Q \ll Q_{\rm eq}$]%
{\small\sl Creation of a dilaton black hole in the high field regime
  with $Q \ll Q_{\rm eq}$. A black hole created at {\rm C} is
  described by a neutral approximation until inflation ends.}
\label{fig-eci3a}
\end{figure}

If, on the other hand, $g'\!\!<\!\!g$ ($Q\!\!<\!\!Q_{\rm
  eq}$), then the Maxwell term will at first be negligible. The first
stage of the black hole evolution will thus be described by the
neutral approximation, with both fields decreasing. Then there are two
possibilities. The first is that $\phi$ eventually reaches $\phi_{\rm
  eq}$. This would lead to a pseudo-static approximation, followed by
a neutral approximation (see Fig.~\ref{fig-eci3}).
The other possibility is that the fields enter the low field regime
while $\phi$ is still larger than $\phi_{\rm eq}$. In this case the
whole evolution will be described by a neutral approximation (see
Fig.~\ref{fig-eci3a}).

The instanton action will now be
\begin{equation}
  I_{S^1 \times S^2}^{\rm Re} = - \frac{\pi}{\tilde{H}_0^2 (1+g')} = -
  \frac{\pi}{3 H_0^2 (1+g')},
\label{eq-action-bh3}
\end{equation}
and the pair creation rate is given by:
\begin{equation}
  \Gamma_{\rm other} = \exp \left( - \frac{\pi}{3 H_0^2}\, \frac{1 +
    3 g'}{1+g'} \right).
\label{eq-pcr-other}
\end{equation}
Since both $ \tilde{g} $ and $g'$ must be positive and less than $1$,
these results do not differ qualitatively from the pseudo-static case.
In particular, the upper bound for the charge, corresponding to
$g'=1$, decreases as the fields roll down, so that the black holes of
highest charge are created early in inflation. When this upper bound
becomes less than $1$, only neutral black holes can be created, since
the charge is quantised.

\subsection{Low Field Regime}

\subsubsection*{Neutral Creation}

\begin{figure}[htb]
  \hspace*{\fill} \vbox{\epsfxsize=.5\textwidth
  \epsfbox{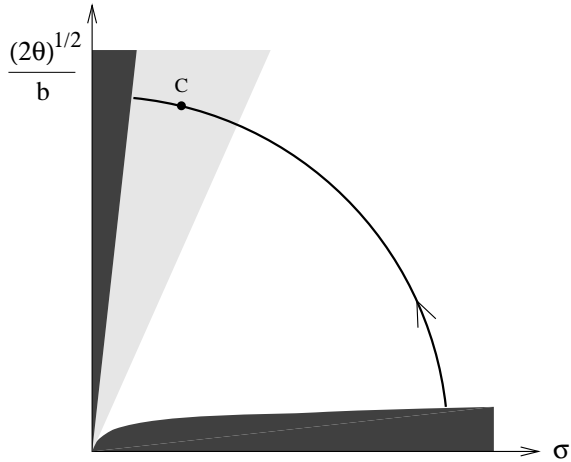}} \hspace*{\fill}
\caption[Creation of a dilaton black hole in the low field regime with
$Q<Q_{\rm eq}$]%
{\small\sl Creation of a dilaton black hole in the low field regime
  with $Q<Q_{\rm eq}$. A black hole created at {\rm C} is described by
  a neutral approximation until inflation ends.}
\label{fig-eci4}
\end{figure}

For the black hole solutions in the low field regime, we found the
neutral approximation, Eqs.~(\ref{eq-neutral-sigI}) --
(\ref{eq-neutral-beta}), by neglecting the Maxwell term. If such
black holes are created via an instanton, the black hole charge must
be small enough to justify this assumption: $g' \approx 0$, or
equivalently,
\begin{equation}
Q \ll Q_{\rm eq},
\end{equation}
where $Q_{\rm eq}$ is given by Eq.~(\ref{eq-qeq}). The process is then
very simple, as shown in Fig.~\ref{fig-eci4}.
After the beginning of inflation, by
Eq.~(\ref{eq-zeta}), the fields move counterclockwise along a circle
around the origin. By Eq.~(\ref{eq-zeta3}), when the pair creation
takes place, the fields continue to move along the original circle,
until inflation ends. The Euclidean action and pair creation rate are
given by Eqs.~(\ref{eq-action-bh3}) and (\ref{eq-pcr-other}), with
$g'$ set to zero.

\subsubsection*{Higher Charges}

\begin{figure}[htb]
  \hspace*{\fill} \vbox{\epsfxsize=.5\textwidth
  \epsfbox{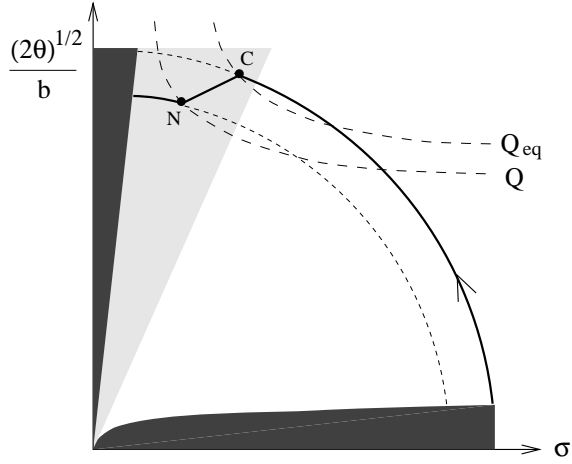}} \hspace*{\fill}
\caption[Creation of a dilaton black hole in the low field regime with
$Q>Q_{\rm eq}$]%
{\small\sl Creation of a dilaton black hole in the low field regime
  with $Q>Q_{\rm eq}$. Black hole pair creation takes place at {\rm
    C}. After the dilaton field has increased sufficiently to
  compensate for the high charge, the black hole enters a neutral
  phase.}
\label{fig-eci5}
\end{figure}

Black holes created in the low field regime with a non-negligible
charge, $0 < Q \leq Q_{\rm eq}$, quickly evolve towards the neutral
approximation, since $\phi_{\rm eq}$ will decrease faster than $\phi$.
Even if the charge is nearly maximal, $Q > Q_{\rm eq}$, the dilaton
will only increase until $\phi = \phi_{\rm eq}$, and then approach the
neutral approximation. The corresponding path in phase space is shown
in Fig.~\ref{fig-eci5}.
 The Euclidean action and pair creation rate are given by
Eqs.~(\ref{eq-action-bh3}) and (\ref{eq-pcr-other}).

\sect{Summary} \label{sec-sum}

We found instantons describing the pair creation of dilatonic black
holes during extended chaotic inflation.  Working in the Einstein
frame, we presented Euclidean solutions for the background and black
hole spacetimes that satisfy the no boundary proposal. We determined
the complex initial values of the dilaton and inflaton. From the
Euclidean action we calculated the pair creation rate.

There is no dilatonic de~Sitter solution with a fixed cosmological
constant. An inflationary universe, with a slowly decreasing effective
cosmological constant, provides the simplest background for the
creation of cosmological dilaton black holes. Such black holes are
described by Charged Nariai-type solutions, which are given by the
topological product of 1+1-dimensional de~Sitter space with a round
two-sphere.  They were presented recently in the context of a fixed
cosmological constant. Adjusting for the time dependence of the
effective cosmological constant, we found two types of approximate
black hole solutions, which are attractive in different regions of the
dilaton-inflaton phase space.

In the pseudo-static approximation, the dilaton evolves such that it
balances the Maxwell and the cosmological terms in the dilaton
equation of motion.  In the neutral approximation, the black hole
charge is small, so that the Maxwell term can be neglected; the
spatial topology is non-trivial, but the evolution of the dilaton
field is the same as in inflation without a black hole.  While the
inflaton and dilaton fields are still large, i.e.\ as long as
Eq.~(\ref{eq-high}) holds, the black hole solutions are attracted to
the pseudo-static approximation. Later the Maxwell term rapidly
becomes negligible, and the black hole solutions are attracted by the
neutral approximation.

We found that dilatonic black holes are plentifully produced near the
Planck era. They become highly suppressed as the cosmological constant
decreases, similar to the non-dilatonic case. Due to the
time-dependence of the Maxwell coupling, however, there is one
significant qualitative difference to black hole pair creation in
standard inflation: the highest possible black hole charge is maximal
at the beginning of inflation, and decreases during the roll-down of
the inflaton field. In the late stages of inflation (the low field
regime) it increases again.

We have considered Charged Nariai black holes because they are the
only RNdS solutions of standard Einstein-Maxwell theory that have an
analogue in dilatonic theories with a Liouville
term~\cite{PolWil94,Bou96}.  In standard inflation, however, most
black holes are nucleated via a different instanton, the ``lukewarm''
solution, which has no exact dilatonic analogue.  In order to obtain
an estimate of the number of magnetically charged dilaton black holes
present in the universe today, it will be necessary to search for an
approximate dilatonic equivalent of the lukewarm solution. We hope to
make some progress on this question.

\end{document}